\documentclass{article}

\author{Alexis Darrasse}
\title{Random XML sampling the Boltzmann way}

\begin{document}
\maketitle

In this article we present the prototype of a framework capable of producing,
with linear complexity, uniformly random XML documents with respect to a given
RELAX NG grammar. The generation relies on powerful combinatorial methods
together with numerical and symbolic resolution of polynomial systems.

\section{Introduction}

The Extensible Markup Language (XML) is extensively used today, either to
encode documents (like in XHTML) or to serialize structured data. The XML
standard\cite{XML} only defines some basic syntax rules followed by well-formed
documents. However applications often define a set of higher order
syntactic\footnote{often called semantic} rules that an XML document must
respect to be considered as valid for the given application.  A set of such
rules is called a schema and is defined in one of several languages, like DTD,
XML Schema, RELAX NG and others. We chose to deal with RELAX NG for its
simplicity and solid theoretical basis.

Our work is based on the observation that RELAX NG\cite{RELAXNG} has
essentially the same expressive power as the specifications of combinatorial
structures using only union, product, sequence and allowing for recursive
definitions. This means that valid XML documents are just trees and can be
efficiently generated using a Boltzmann sampler\cite{DuFlLoSc04}.

In Boltzmann sampling we can easily derive a generator from the description of
a combinatorial class. This generator has the following characteristics:
\begin{itemize}
  \item it needs a precalculation that must be done once and involves finding a
  particular solution of a polynomial system,
  \item the sampling itself involves only basic mathematic operations and has a
  linear complexity in the size of the generated object and
  \item the size of the generated object is a random variable following a power
  law distribution.
\end{itemize}

We go from a RELAX NG grammar to a random XML document in three steps:
\begin{enumerate}
  \item translating the grammar to a system of equations,
  \item solving the system of equations and
  \item sampling XML documents.
\end{enumerate}

The first two steps need to be executed only once for a given grammar and their
cost is only dependant on the complexity of the grammar. The final step has a
complexity that is linear in the size of the result.

The current prototype includes a ruby program executing the first and last step
and a maple program dealing with the second step. Future versions could consist
of a single program not depending on a computer algebra system.

Even in its current form, our framework is capable of generating documents for
all RELAX NG grammars that we were able to find, include XHTML, MathML, SVG,
DocBook, OpenDocument and RELAX NG itself.

\section{Translating the RELAX NG}

We parse the RELAX NG document to get a combinatorial description of the
grammar in the form of an Abstract Syntax Tree (AST).  Every RELAX NG element
is matched to a combinatorial construction, for example \texttt{choice} to
union, \texttt{group} to product, \texttt{oneOrMore} to sequence etc.

The mapping from RELAX NG to combinatorial constructions is not unique, we thus
have to make a few arbitrary decisions. First of all, we must decide of a way to
count the size of an XML document, while satisfying the constraint that there
must be a finite number of XML documents for a given size. Our choice is to
count the number of elements plus the number of attributes.

Some facts on the AST: at the root we have the definition of \texttt{start}
which is our entry point to the grammar and the definition of every
\texttt{element} of the grammar. We can thus represent the AST as a forest, with
every root being the definition of an \texttt{element}. \texttt{ref} elements
found in different parts of the AST point to one of these definitions, which can
lead us to see the AST also as a graph.

\subsection*{Note on \texttt{data}}

Our framework concentrates on the tree-form structure of the XML document and
treats \texttt{data} elements as simple leaves. These elements however
represent data respecting arbitrarily complex datatypes which cannot be
reasonably treated by a generic framework. We thus provide the possibility to
call an arbitrary (written in Ruby since performed at step 3) function when the
generator needs a value of a \texttt{data} element.

Even though the RELAX NG language does not provide a collection of basic
datatypes, the XML Schema Datatypes\cite{XSD} are used as such. We thus
provide default samplers for each of the datatypes contained in this
library, while preserving the possibility for the users to overload them.

\subsection*{Note on uniformity}

One of the main advantages of using Boltzmann sampling is that it guarantees the
uniformity of the distribution inside each size class. The simplifications made
in our prototype break this uniformity in two ways:
\begin{itemize}\itemsep=0pt\topsep=0pt\partopsep=0pt\parskip=0pt\parsep=0pt
  \item sets of attributes are considered as sequences, so \texttt{<a x="foo"
  y="bar"/>} and \texttt{<a y="bar" x="foo"/>} count as two different XML
  documents, while they are the same (but two attributes of the same element
  will always have different names) and
  \item the \texttt{interleave} element, which calls for the interleaving of
  it's arguments, is not taken into account, so all the interleavings count as a
  single element.
\end{itemize}

\section{Solving the system of equations}

The AST is fed to a Maple program that solves the set of equations defining the
grammar and provides us with the constants needed for sampling. The calculation
time needed for this operation depends heavily on the size and structure of the
grammar. The user can use this Maple program as a black box, as it is not
interactive.  But for the purpose of explanation we now describe it a little
more precisely.

A RELAX NG grammar defines, in our point of view, a (combinatorial) family of
trees. As for every other combinatorial class, a generating function $C(x) =
\sum_{n=0}^\infty t_n x^n$ is assocatied to this family, with $t_n$ being the
number of different trees of size $n$. An important parameter is the radius of
convergence $\rho$ of the generating function $C(x)$. The Boltzmann sampler
uses the value of $C(x)$ and some related generating functions for a given
parameter $x$. In the case of trees, $\rho$ is an excellent choice for
$x$\cite{DuFlLoSc04}.

The AST of the grammar can be directly translated into an algebraic system
(that can itself be simplified to a polynomial one) defining the generating
function $C(x)$. To be able to sample we need to solve two problems: evaluate
the radius of convergence of the system and find its only solution that
corresponds to $C$.

A major difficulty comes from the fact that simply solving the system for a
given $x$ gives us a set of solutions with no simple way of finding the good
one. Thankfully, a combinatorial interpretation of Newton's
method\cite{PiSaSo08} gives us a very efficient and guaranteed numerical
algorithm for evaluating $C(x)$.

As for the radius of convergence, for the time being, using a dichotomy
approach and Newton algorithm we obtain an estimation. We can also (see
section~\ref{sec:wip}) take advantage of the particularities of the generating
functions arising from grammars defining trees, in which the radius of
convergence can be automatically calculated, as explained in \cite{DuFlLoSc04}.

To experiment the solving algorithm, we tried it on different RELAX NG grammars
(found for most of them on the internet). Table~\ref{table:eval} shows some
measures on the grammars related to their complexity, together with the time
needed to evaluate the corresponding generating functions. The size of the
corresponding polynomial systems is also mentioned, showing that solving them
with a symbolic approach is a challenge. Here is a precise description of the
column headers:

\begin{description}\itemsep=0pt\topsep=0pt\partopsep=0pt\parskip=0pt\parsep=0pt
  \small
  \item[sing.] Lower bound on the singularity of the system of equations.
  \item[file.] Size of the file containing the grammar in Simple RELAX NG.
  \item[\# el.] Number of \texttt{element} definitions.
  \item[s.c.c.] Size of the biggest strongly connected component of the grammar.
  \item[newton] Time needed to evaluate the values of the generating functions
  using the newton algorithm.
  \item[\# eq.] Number of equations defining the polynomial system.
  \item[\# mon.] Number of monomials in the polynomial system.
  \item[\# sol.] Number of solutions of the zero-dimensional system, given a
  parameter smaller than the singularity.
\end{description}

\begin{table}
  \setlength{\tabcolsep}{2pt}
  \centering
  \scriptsize
  \begin{tabular}{| c | c | c | c | c || c || c | c | c |}
    \hline
    grammar       & file & sing.   & \# el. & s.c.c. & newton   & \# eq. & \# mon. & \# sol.\\ \hline
    ternary trees & 1024 & 0.52912 & 2      & 1      & 0.260s   & 3      & 10      & 3      \\
    RSS           & 9.5K & 0.44721 & 10     & 1      & 0.320s   & 16     & 79      & 2      \\
    PNML          & 23K  & 0.22526 & 22     & 1      & 0.322s   & 36     & 193     & 4      \\
    ILP 1         & 21K  & 0.23696 & 20     & 6      & 0.416s   & 27     & 211     & 9      \\
    ILP 6         & 99K  & 0.13951 & 51     & 31     & 0.828s   & 72     & 948     & 6      \\
    RELAX NG      & 124K & 0.04127 & 33     & 18     & 0.696s   & 114    & 3725    & 32     \\
    XSLT          & 168K & 0.05283 & 40     & 17     & 0.469s   & 122    & 1503    & 10     \\
    XHTML         & 289K & 0.02456 & 47     & 32     & 0.932s   & 134    & 2077    & 26     \\
    XML Schema    & 237K & 0.08615 & 59     & 9      & 0.528s   & 188    & 143465  &        \\
    XHTML basic   & 284K & 0.03039 & 53     & 38     & 1.080s   & 96     & 2073    & 13     \\
    XHTML strict  & 1.2M & 0.01991 & 80     & 58     & 2.414s   & 151    & 6445    & 32     \\
    XHTML         & 1.5M & 0.01609 & 93     & 66     & 3.798s   & 172    & 8449    & 56     \\
    SVG tiny      & 1.6M & 0.03542 & 49     & 7      & 0.371s   & 101    & 4390    & 34     \\
    SVG full      & 6.3M & 0.01834 & 118    & 27     & 0.718s   & 232    & 13831   &        \\
    MathML        & 2.2M & 0.00318 & 182    & 48     & 2.432s   & 265    & 265159  & 18     \\
    OpenDocument  & 2.8M & 0.01757 & 500    & 101    & 6.544s   & 814    & 8890517 &        \\
    DocBook       & 11M  & 0.01627 & 407    & 295    & 143.411s & 977    & 183051  &        \\
    \hline
  \end{tabular}
  \caption{Evaluating the generating functions of different RELAX NG grammars.
  Measures made on a 3.2GHz Xeon with 6GB of RAM, using Maple 10 in a 64bit
  environment.}
  \label{table:eval}
\end{table}

These results show that with the numerical method we were able to deal with all the
grammars in a very reasonable time.

\section{Generating XML documents}

To generate the XML documents the Boltzmann way, we explore the AST, starting
from the \texttt{start} element of the grammar and treating the different
elements the following way:
\begin{itemize}\itemsep=0pt\topsep=0pt\partopsep=0pt\parskip=0pt\parsep=0pt
  \item for a union we choose randomly between the two siblings,
  \item for a product we take both children,
  \item for a sequence we take the child a random number of times,
  \item for a \texttt{ref} we continue at the corresponding \texttt{element},
  \item for a leaf produce the corresponding value.
\end{itemize}

Thus the cost to produce an XML document is proportional to the size of the
document and the constant factor depends on the grammar. That size however is
random and actually follows a power law distribution. This means that we
generate a very large number of very small documents and from time to time a
huge document, eventually bigger than we can handle.

For this reason we include a mechanism for rejecting documents whose size falls
outside a given window. The cost of the rejection can be precisely
estimated\cite{DuFlLoSc04}: generating a document of size $n(1\pm\epsilon)$,
for a fixed tolerance $\epsilon$, costs $O(n)$, while generating a document of
size exactly $n$ costs $O(n^2)$ (costs are average, in the worst case the
generator never returns a valid document, but this happens with probability
zero).

\section{Work in progress}
\label{sec:wip}

The current state of our work shows what can be achieved by applying the
Boltzmann sampling techniques in the problem of sampling random XML documents.
We are in the process of completing and extending our framework in order to
create tools useful as-is in a large number of use cases.

\paragraph{Efficient implementation and integration of the framework.}
The current implementation of our framework is a prototype with a design
targeting flexibility and ease of development. Consequently it lacks in
efficiency and ease of deployment, which will be the goals of future versions.

\paragraph{Progressive serialisation of the generated document.}
To be able to sample documents of millions of elements, or more, we can no
longer keep the whole document in memory. The solution is to write the data on
disk as soon as we know its position in the resulting document. This should be
trivial for grammars not containing \texttt{interleave} constructions and could
be reasonably treated even in that case.

\paragraph{Conserve the entropy, guarantee the precision.}
By implementing the non-deterministic parts of the samplers using bitwise
comparisons, we can make sure not to use more entropy bits than necessary. At
the same time we can check that the precision used for calculating the constants
is sufficient, and if not, get more precision from the solver.

\paragraph{Assure uniformity in all cases.}
The bias introduced in the sampling by the simplified treatment of the
\texttt{interleave} element and the attributes could be dealt with, if
necessary.  The set construction is already treated in the Boltzmann sampling,
but leads to systems that are no longer polynomial. The treatment of the
interleave construction is work in progress.

\bibliographystyle{plain}
\bibliography{xmlgen}

\end{document}